\documentclass[aps,pra,superscriptaddress,amsmath,amssymb,showpacs]{revtex4-2}
\usepackage[english]{babel}
\usepackage[utf8]{inputenc}
\usepackage{amssymb,bm}
\usepackage{graphicx}
\usepackage{amsmath}
\usepackage{color}
\usepackage{comment}
\usepackage[colorlinks=true,citecolor=blue,urlcolor=blue]{hyperref}
\def\DsFE{D_{s1}(2460)}
\def\DsTS{D_{s1}(2536)}

\begin{document}

	\title{Phenomenology of  $D_{s1}$ mesons radiative transitions}
	
\author{A.~E.~Bondar}\email{A.E.Bondar@inp.nsk.su}
\author{A.~I.~Milstein}\email{A.I.Milstein@inp.nsk.su}
\affiliation{Budker Institute of Nuclear Physics of SB RAS, 630090 Novosibirsk, Russia}
\affiliation{Novosibirsk State University, 630090 Novosibirsk, Russia}

	\date{\today}

	\begin{abstract}
		We discuss radiative transitions $D_{s1}\rightarrow D_s\gamma$ and $D_{s1}\rightarrow D_s^*\gamma$ of $p$-wave mesons. Since $D_{s1}$ mesons  have no a certain $C$-parity, and the masses of quarks in these mesons differ significantly, then due to the spin-orbit interaction each of  $p$-wave mesons with the total angular momentum $J=1$ is a superposition of states with the total spin of quarks $S_{tot}=0$ and $S_{tot}=1$. We explain why the partial width of $D_{s1}(2460)\to D_s\gamma$ radiative transition is significantly larger than the corresponding value for $D_{s1}(2536)\to D_s\gamma$. We also predict the corresponding partial widths of $B_{s1}$ and $B_{c1}$.
		\end{abstract}
	
	\maketitle
	
	\section { Introduction}
	Currently, various phenomenological approaches are used to study spectroscopy and radiative transitions of hadronic states containing heavy quarks. In particular, one can try to use a nonrelativistic quark model with account for the first relativistic corrections. The corrections are especially important if the photon radiation amplitude obtained in the leading nonrelativistic approximation is suppressed for one reason or another. In this case, when calculating the radiation amplitudes, in addition to relativistic corrections it is also necessary to take into account the retardation effect in the photon wave function. A similar problem arises in both atomic and nuclear physics. An example is a single-photon transition $2s\to 1s$ in the hydrogen atom \cite{Drake:1971zz} and bremsstrahlung during scattering of two identical nonrelativistic nuclei \cite{Obraztsov:2021tip}. This is precisely the situation that arises when studying radiative transitions of $p$-wave states, for example,  $c\bar s$ systems with the total momentum $J=1$. The nonrelativistic electric dipole moment of this system is 
		\begin{align} \label{dnr}
		&\bm d=\left[ \dfrac{e_{\bar s}}{m_s}-\dfrac{e_{c}}{m_c}\right]\,M_R\bm r\approx \dfrac{e_{\bar s}}{3m_s}\,M_R\bm r\,,
	\end{align}
	where $m_s$ and $e_{\bar s}=e/3$ are the mass and charge of  $\bar s$ quark, $m_c$ and $e_{c}=2e/3$ are the mass and charge of $c$ quark, $e$ is the proton charge, $\bm r=\bm r_{\bar s}-\bm r_{c}$, $M_R=m_cm_s/(m_c+m_s)$ is the reduced mass. In the center-of-mass frame $\bm r_{\bar s}=M_R\bm r/m_s$ and $\bm r_{c}=-M_R\bm r/m_c$. For the estimate,  the values $m_s= 0.5\,\mbox{GeV}$ and $m_c= 1.7\,\mbox{GeV}\,$ are used.  A strong compensation in the magnitude of  electric dipole moment is seen.
	  
	  The nonrelativistic dipole moment \eqref{dnr} is independent of  quark spins  $\bm S_{\bar s }$ and  $\bm S_c$ and cannot lead to radiative transitions between the states with   $S_{tot}=1$  and $S_{tot}=0$ of the total spin  $\bm S_{tot}=\bm S_{\bar s} +\bm S_{c }$.  Meanwhile, there are two $p$-wave mesons consisting of $c$ and $\bar s$ quarks, $D_{s1}(2460)$ and $D_{s1}(2536)$, with the total momentum $J=1$ and close masses. Each of these mesons is a superposition of states with $S_{tot}=1$ and $S_{tot}=0$ \cite{Godfrey:1985xj,Barnes:2005pb,Matsuki:2010zy,Song:2015nia}. This is because the fine structure operator of the system depends not only on $\bm S_{tot}$ but also on the operator
	  	$\bm \Sigma=\bm S_{\bar s}-\bm S_{c }$ having nonzero matrix elements between the states with different values of the operator $\bm S_{tot}$.
Therefore, the total spin is not conserved, unlike the total angular momentum $J$, corresponding to the operator $\bm J=\bm L +\bm S_{tot }$, where $\bm L$ is the angular momentum operator of the system. In addition, the operator responsible for radiative transitions between states with $L=1$ and $L=0$ also  contains  contributions proportional to $\bm S_{tot}$ and $\bm \Sigma$. This, as it is shown in our paper, explains the unusual ratio of widths of   $D_{s1}$-meson radiative transitions \cite{Bondar:2025gsh}. The conclusions of Ref.~\cite{Bondar:2025gsh} are based on the analysis of experimental data of the $BaBar$ collaboration \cite{BaBar:2006eep}.
As noted in Ref.~\cite{Bondar:2025gsh}, the width of  transition $D_{s1}(2536)\to D_s\gamma$ is less than $8\,\mbox{KeV}$, and the width of transition $D_{s1}(2460)\to D_s\gamma$ is less than $2.3\,\mbox{KeV}$, which is strikingly different from the generally accepted expectations presented in Table \ref{tb1}. Note that the estimation for the width of $D_{s1}(2460)\to D_s\gamma$ transition in Ref.~\cite{Bondar:2025gsh} was based on the account of the nonrelativistic dipole moment contribution solely.  We show that this assumption is not valid due to importance of relativistic corrections. Taken these corrections into account withing a potential model, we obtain  the formulas for the probabilities of $D_{s1}\rightarrow D_s\gamma$, $D_{s1}\rightarrow D_s^*\gamma$, $B_{s1}\rightarrow B_s\gamma$, $B_{s1}\rightarrow B_s^*\gamma$, $B_{c1}\rightarrow B_c\gamma$, and $B_{c1}\rightarrow B_c^*\gamma$ transitions.

Our work is organized as follows. We first derive the spin-dependent corrections to the operator responsible for radiative transitions, make qualitative predictions for the corresponding probabilities, and compare them with available experimental data. The Conclusion formulates the main results of our work.
	
	\section {Fine structure of $c\bar s$ system in the radiation field}
	To obtain a radiative-transition operator that takes into account relativistic corrections and the retardation effect, we consider the Hamiltonian of $c\bar s$ system  in the radiation field. This Hamiltonian contains several terms which we discuss separately.
	The vector potential $\bm A(\bm r,t)$ of  electromagnetic wave in the final state, the corresponding electric field $\bm{\mathcal E}(\bm r,t)$ and magnetic field $\bm{\mathcal H}(\bm r,t)$ have the form
		\begin{align}\label{AEH}
		&	 \bm A(\bm r,t)=\bm A_W e^{i(\omega t-\bm k\cdot\bm r)}\,,\quad \bm{\mathcal H}(\bm r,t)=-i[\bm k\times\bm A(\bm r,t)]\,,\quad \bm{\mathcal E}(\bm r,t)=-i\omega\bm A(\bm r,t)\,,
	\end{align}
	where $\bm A_W$ is a constant vector perpendicular to the momentum $\bm k$ of  photon, the system of units $\hbar=c=1$ is used.
	
	We start with the Hamiltonian $H_{0}$, which accounts for the first spin-independent relativistic correction to the kinetic energy, 
		\begin{align} \label{nr}
		&H_{0}=\dfrac{\bm\pi_{\bar s}^2}{2m_s}+\dfrac{\bm\pi_{c}^2}{2m_c}+U_{g }(r)+U_{conf }(r)   - \dfrac{(\bm\pi_{\bar s}^2)^2}{8m_s^3}-\dfrac{(\bm\pi_{c}^2)^2}{8m_c^3}\,,\nonumber\\
		&\bm\pi_{\bar s}=\bm p-e_{\bar s}\bm A(\bm r_{\bar s},t)\,,\quad \bm\pi_{c}= -\bm p-e_{c}\bm A(\bm r_{ c},t)\,,\nonumber\\
		&U_{g }(r)=-\dfrac{g}{r}\,,\quad U_{conf }(r)=br\,,\quad g=\dfrac{4}{3}\,\alpha_s\,,
	\end{align} 
where $\bm p=-i\bm\nabla$, $g$ and $b$ are the parameters of our model. The values of these parameters are fixed by comparing the predictions for the masses of states with known experimental values \cite{ParticleDataGroup:2024cfk}. 

The potential energy $U_{g}(r)$ of  $c$ and $\bar s$ interaction is the zero component of a Lorentz vector, and $U_{conf}(r)$ is a Lorentz scalar \cite{Godfrey:1985xj}. This is why $U_{g}(r)$ and $U_{conf}(r)$ enter differently into the spin-dependent relativistic correction $H_{S}$ to the Hamiltonian, 
	\begin{align} \label{rel1}
		&H_{S}=\dfrac{1}{2}\left(\dfrac{g}{r^3}-\dfrac{b}{r}\right)\left(\dfrac{[\bm r\times\bm\pi_{\bar s}]\cdot \bm S_{\bar s}}{m_s^2}-\dfrac{[\bm r\times\bm\pi_{c}]\cdot \bm S_{c }}{m_c^2}\right)\nonumber\\
		&-\dfrac{g}{m_{s}m_c r^3}\Big([\bm r\times\bm\pi_{c}]\cdot \bm S_{\bar s}-[\bm r\times\bm\pi_{\bar s}]\cdot \bm S_{c }\Big) \nonumber\\
		&+\dfrac{g}{m_{s}m_c }\Bigg[\dfrac{3(\bm n\cdot\bm S_{tot})^2-\bm S_{tot}^2}{2r^3}+\dfrac{4\pi}{3}\,\delta(\bm r)\,\left(\bm S_{tot}^2-\dfrac{3}{2}\right)\Bigg]\,,
	\end{align} 
where $\bm n=\bm r/r$. Recall that $\bm S_{tot}^2-3/2=2\bm S_{\bar s}\bm S_c$. Finally, there are two more corrections to the Hamiltonian that do not explicitly include the potentials $U_{g}(r)$ and $U_{conf}(r)$. The first correction $H_{M}$ is related to the interaction of magnetic field with the magnetic moments of  quarks, 
	\begin{align} \label{M}
		&H_{M}=-\dfrac{e_{\bar s}}{m_s}\bm S_{\bar s}\cdot\bm{\mathcal H}(\bm r_{\bar s},t)-\dfrac{e_{c}}{m_c}\bm S_{c}\cdot\bm{\mathcal H}(\bm r_{\bar c},t)\,.
	\end{align}  
The second correction $H_{T}$ is related to the Thomas precession, which occurs due to the acceleration of a particle in the electric field of radiation,
	\begin{align} \label{T}
		&H_{T}=-\dfrac{e_{\bar s}}{4m_s^2}\Big([\bm\pi_{\bar s}\times\bm{\mathcal E}(\bm r_{\bar s},t)]-[\bm{\mathcal E}(\bm r_{\bar s},t)\times\bm\pi_{\bar s}]\Big)\cdot\bm S_{\bar s}\nonumber\\
		&	-\dfrac{e_{c}}{4m_c^2}\Big([\bm\pi_{c}\times\bm{\mathcal E}(\bm r_{c},t)]-[\bm{\mathcal E}(\bm r_{c},t)\times\bm\pi_{c}]\Big)\cdot\bm S_{c}\,.
	\end{align} 
For $U_{conf}(r)=0$, the Hamiltonian of a system  with different masses and charges in a radiation field  can be found in Refs.~\cite{Pilkuhn:1979ps,Lee:2001wwa}. The contribution of $U_{conf}(r)$ to the spin-dependent part of the Hamiltonian is proportional to the constant $b$ and is contained only in the correction $H_{S}$.
	
We now have everything we need to study both the structure of p-wave mesons and their radiative transitions.
	
	\section {Structure of $D_{s1}$ mesons} 
	We set $\bm A(\bm r,t)=0$ and obtain the Hamiltonian describing the fine structure of  $c\bar s$ system with $L=1$,
		\begin{align} \label{s}
		&H=H_{nr}+\Delta H_{nr}+\Delta H_S\,,\nonumber\\
		& H_{nr}=\dfrac{\bm p^2}{2M_R}-\dfrac{g}{r}+br\,,\quad \Delta H_{nr}=-\dfrac{(\bm p^2)^2}{8m_s^3} -\dfrac{(\bm p^2)^2}{8m_c^3}\,,\nonumber\\
		&\Delta H_S=\dfrac{1}{4}\left(\dfrac{g}{r^3}-\dfrac{b}{r}\right)\,\bm L\cdot\left[\left(\dfrac{1}{m_s^2}+\dfrac{1}{m_c^2}\right)\bm S_{tot}+\left(\dfrac{1}{m_s^2}-\dfrac{1}{m_c^2}\right)\bm\Sigma\right]\nonumber\\
		&+\dfrac{g}{m_{s}m_c r^3}\bm L\cdot \bm S_{tot}+\dfrac{g}{2m_{s}m_c r^3}\Big[3(\bm n\cdot\bm S_{tot})^2-\bm S_{tot}^2\Big]\,.
	\end{align} 
Recall that $\bm p^2=p_r^2+\bm L^2/r^2$, where $-p_r^2$ is  the radial part of the Laplacian. The Hamiltonian \eqref{s} coincides with the Hamiltonian in the model \cite{Godfrey:1985xj} with  one difference,  in our model only the first correction $\Delta H_{nr}$ to the kinetic energy is taken into account. We have checked that such a replacement does not significantly affect the numerical results, but essentially  simplifies the calculations.

The eigenfunctions of the Hamiltonian with $J=1$, $L=1$ and $J_z=\mu$, $\Psi_1$ and $\Psi_2$, are a linear combination of the functions $\psi_1$ and $\psi_2$,
	\begin{align} \label{psi12}
		&\Psi_1=c_1\psi_1+c_2\psi_2\,,\quad \Psi_2=c_2\psi_1-c_1\psi_2\,,\nonumber\\
		&\psi_1=Y_{1\,\mu}(\theta,\varphi)\,|S=0,S_z=0\rangle\,R_1(r)\,,  \nonumber\\
		& \psi_2=\sum_{\mu'}C^{1\mu}_{1\mu',1 (\mu-\mu')}\,Y_{1\,\mu'}(\theta,\varphi)|S=1,S_z=\mu-\mu'\rangle\,R_1(r)\,,
	\end{align} 
where $Y_{1\,\mu}(\theta,\varphi)$ is the spherical function, $C^{1\mu}_{1\mu',1 (\mu-\mu')}$ are the Clebsch-Gordan coefficients,
$R_1(r)$ is the radial wave function with the angular momentum $L=1$. The corresponding energy and the coefficients $c_{1,2}$ are found from the solution of the secular equation \cite{LLQM}. 

The matrix elements of the operator $\Delta H_S$ read 
	\begin{align} \label{DelH}
		&(\Delta H_S)_{11}=0\,,\quad (\Delta H_S)_{22}=  -\dfrac{1}{4}\left(\dfrac{1}{m_s^2}+\dfrac{1}{m_c^2}\right)\left(\left\langle\dfrac{g}{r^3}\right\rangle-
		\left\langle\dfrac{b}{r}\right\rangle\right)
		-\dfrac{1}{2m_{s}m_c}\left\langle\dfrac{g}{r^3}\right\rangle\,,\nonumber\\
		&(\Delta H_S)_{21}=(\Delta H_S)_{12}=\dfrac{1}{2\sqrt{2}}\left(\dfrac{1}{m_s^2}-\dfrac{1}{m_c^2}\right)\left(\left\langle\dfrac{g}{r^3}\right\rangle-\left\langle\dfrac{b}{r}\right\rangle\right)\,.
	\end{align} 
Let $\Psi_1$ corresponds to the lower energy $E_1=E_1^{(0)}+\Delta E_1$ with $\Delta E_1<0$, and $\Psi_2$ to the higher energy $E_2=E_1^{(0)}+\Delta E_2$ with $E_2>0$, where $E_1^{(0)}$ is the eigenvalue of the Hamiltonian $H_{nr}$. The state $\Psi_1$ corresponds to $D_{s1}(2460)$ meson, and $\Psi_2$ to $D_{s1}(2536)$ meson. Solving the secular equation, we obtain 
	\begin{align} \label{c12}
		&\Delta E_1=\dfrac{(\Delta H_S)_{22}}{2}-\gamma\,,\quad \Delta E_2=\dfrac{(\Delta H_S)_{22}}{2}+\gamma\,,\quad \gamma=\sqrt{\dfrac{(\Delta H_S)_{22}^2}{4}+(\Delta H_S)_{12}^2}\,,\nonumber\\
		& c_1=-\mbox{sgn}(\Delta H_S)_{12}\,\sqrt{\dfrac{\Delta E_2}{2\gamma}}\,,\quad  c_2=\sqrt{\dfrac{|\Delta E_1|}{2\gamma}}\,.
	\end{align} 
To calculate the matrix elements of radiative transitions, we need to know the wave functions not only of $D_{s1}$ mesons, but also of $D_s$ mesons with $L=0$, $S_{tot}=0$ and $J=0$. We use the wave functions calculated by means of
 the variational principle \cite{LLQM}. For states with $L=0$, we write the radial wave function $R_0(r)$ as
	\begin{align} \label{WF0}
		&R_0(r)=\sqrt{N_0}\,\exp(-\omega_0^2r^2/2)\,,\quad N_0=\dfrac{4}{\sqrt{\pi}}\,\omega_0^3  \,,
	\end{align}
	where $\omega_0$ is a variational parameter. Then
	\begin{align}\label{E00}	
		&E_0^{(0)}=\dfrac{3\omega_0^2}{4M_R}+\dfrac{2}{\sqrt{\pi}}\left(\dfrac{b}{\omega_0}-g\omega_0\right)\,,\nonumber\\
		&\dfrac{\partial}{\partial\omega_0}E_0^{(0)}= \dfrac{3\omega_0}{2M_R}-\dfrac{2}{\sqrt{\pi}}\left(\dfrac{b}{\omega_0^2}+g\right)=0\,.
	\end{align} 
Here $E_0^{(0)}$ is the average value of the Hamiltonian $H_{nr}$ over the wave function \eqref{WF0} at $L=0$. Similarly, for $L=1$ the variational radial wave function $R_1(r)$ has the form
	\begin{align} \label{WF1}
		&R_1(r)=\sqrt{N_1}\,r\exp(-\omega_1^2r^2/2)\,,\quad N_1=\dfrac{8}{3\sqrt{\pi}}\,\omega_1^5  \,,
	\end{align}
where the variational parameter $\omega_1$ is found from the relations
	\begin{align}\label{E10}
		&E_1^{(0)}=\dfrac{5\omega_1^2}{4M_R}+\dfrac{4}{3\sqrt{\pi}}\left(\dfrac{2b}{\omega_1}-g\omega_1\right)\,,\nonumber\\
		&\dfrac{\partial}{\partial\omega_1}E_1^{(0)}=\dfrac{5\omega_1}{2M_R}-\dfrac{4}{3\sqrt{\pi}}\left(\dfrac{2b}{\omega_1^2}+g\right)=0\,.
	\end{align}
In these formulas $\omega_0$ and $\omega_1$ are variational parameters that are found by minimizing the corresponding energies. For the average values in the matrix elements, that determine the mixing coefficients of the functions  $\psi_1$ and $\psi_2$, we obtain
	\begin{align} \label{rr3}
		&\int_0^\infty \dfrac{R_1^2(r)}{r^3}\,dr=\dfrac{4}{3\sqrt{\pi}}\,\omega_1^3\,,\quad  
		\int_0^\infty \dfrac{R_1^2(r)}{r}\,dr=\dfrac{4}{3\sqrt{\pi}}\,\omega_1\,,
	\end{align} 
	As a result, for the matrix elements we have
		\begin{align} \label{H12H22}
		&(\Delta H_S)_{11}=0\,,\quad
		(\Delta H_S)_{22}=\dfrac{1}{3\sqrt{\pi}}\left[(b\omega_1-g\omega_1^3)\left(\dfrac{1}{m_s^2}+\dfrac{1}{m_c^2}\right)-\dfrac{2g}{m_sm_c}\omega_1^3\right]\,,\nonumber\\
		&(\Delta H_S)_{12}=(\Delta H_S)_{21}=-\dfrac{\sqrt{2}}{3\sqrt{\pi}}(b\omega_1-g\omega_1^3)\left(\dfrac{1}{m_s^2}-\dfrac{1}{m_c^2}\right)\,.
	\end{align}
Note that the spin-independent relativistic corrections, 
		\begin{align} \label{delnr1}
		&(\Delta H_{nr})_{11}=(\Delta H_{nr})_{22}=-\dfrac{35 \omega_1^4}{32 }\left(\dfrac{1}{m_s^3}+\dfrac{1}{m_c^3}\right)\,,
	\end{align}
do not contribute to the energy difference $E_2-E_1$.
The corresponding  shift of energy for the state with $L=0$ is
	\begin{align} \label{delnr0}
		&(\Delta H_{nr})_{00}=-\dfrac{15 \omega_0^4}{32 }\left(\dfrac{1}{m_s^3}+\dfrac{1}{m_c^3}\right)\,.
	\end{align}
In addition, for $L=0$ it is necessary to take into account the correction
	\begin{align} \label{delS0}
		&(\Delta H_{S})_{00}=-\dfrac{2g\, \omega_0^3 }{\sqrt{\pi}\, m_sm_c}\,.
	\end{align}
	These shifts must be taken into account when comparing the prediction for  energies of  transitions with the experimental values.
	
	\section {Radiative transitions of $D_{s1}$ mesons} 
	To derive the Hamiltonian $H_{rad}$ corresponding to the emission of one photon, it is necessary to keep in $H_{0}$, $H_{S}$, $H_M$ and $H_T$ the terms  linear in $\bm A(\bm r,t)$. Let us represent $H_{rad}$ as follows
		\begin{align} \label{Hrad}
		& H_{rad}=G_0+\dfrac{1}{2}\bm S_{tot}\cdot\bm G_{tot}+\dfrac{1}{2}\bm \Sigma\cdot\bm G_{\Sigma}\,,
	\end{align} 
	where
	\begin{align} \label{G}
		& G_0= -(\bm A_W\cdot\bm p)F_1+\frac{1}{4}\,\{\bm p^2,\,F_3\} (\bm A_W\cdot\bm p)\,,\nonumber\\
		&\bm G_{tot}=-[\bm r\times\bm A_W]\left[\dfrac{1}{2}\left(\dfrac{g}{r^3}-\dfrac{b}{r}\right)f_2
		+\dfrac{g}{m_{s}m_c r^3}f_0\right]\nonumber\\
		&+i[\bm k\times\bm A_W]f_1+\dfrac{i\omega}{4}\Big\{[\bm p\times\bm A_W],f_2\Big\}\,,\nonumber\\
		&\bm G_{\Sigma}=[\bm r\times\bm A_W]\left[-\dfrac{1}{2}\left(\dfrac{g}{r^3}-\dfrac{b}{r}\right)F_2
		+\dfrac{g}{m_{s}m_c r^3}F_0\right]\nonumber\\
		&+i[\bm k\times\bm A_W]F_1+\dfrac{i\omega}{4}\Big\{[\bm p\times\bm A_W],F_2\Big\}\,.
	\end{align} 
	Here
	\begin{align} \label{def}
		&f_0=e_{\bar s}\, e^{-i\bm q_{\bar s}\cdot\bm r}-e_{c}\, e^{i\bm q_{c}\cdot\bm r}\,,\quad
		f_1=\dfrac{e_{\bar s}}{m_s}\, e^{-i\bm q_{\bar s}\cdot\bm r}+\dfrac{e_{c}}{m_c}\, e^{i\bm q_{c}\cdot\bm r}\,,\nonumber\\
		&f_2=\dfrac{e_{\bar s}}{m_s^2}\, e^{-i\bm q_{\bar s}\cdot\bm r}-\dfrac{e_{c}}{m_c^2}\, e^{i\bm q_{c}\cdot\bm r}\,,\quad F_0=e_{\bar s}\, e^{-i\bm q_{\bar s}\cdot\bm r}+e_{c}\, e^{i\bm q_{c}\cdot\bm r}\,,\nonumber\\
		&F_1=\dfrac{e_{\bar s}}{m_s}\, e^{-i\bm q_{\bar s}\cdot\bm r}-\dfrac{e_{c}}{m_c}\, e^{i\bm q_{c}\cdot\bm r}\,,\quad F_2=\dfrac{e_{\bar s}}{m_s^2}\, e^{-i\bm q_{\bar s}\cdot\bm r}+\dfrac{e_{c}}{m_c^2}\, e^{i\bm q_{c}\cdot\bm r}\,,
		\nonumber\\
		&F_3=\frac{e_{\bar s}}{m_s^3}\,e^{-i\bm q_{\bar s}\cdot\bm r}-\frac{e_{c}}{m_c^3}\,e^{i\bm q_{ c}\cdot\bm r}\,,\quad
        \bm q_{\bar s}=\dfrac{M_R}{m_s}\bm k\,,\quad \bm q_{c}=\dfrac{M_R}{m_c}\bm k\,,
	\end{align} 
    and $\{A,B\}=AB+BA\,$.
	\subsection {Radiative transitions of $D_{s1}$ mesons to $D_{s}\gamma$} 
The state of $D_s$ meson is described by the wave function $\psi_0$,
	\begin{align}
		\psi_0=Y_{0\,0}\,|S=0,S_z=0\rangle\,R_0(r)\,.
	\end{align}
To calculate the amplitudes of radiative transitions from the states $\Psi_{1,2}$ to the state $\psi_0$, it is necessary to calculate two matrix elements,
	\begin{align}\label{T1T2}
		T_1=\langle \psi_0|H_{rad}|\psi_1\rangle\,, \quad  T_2=\langle \psi_0|H_{rad}|\psi_2\rangle\,.
	\end{align}	 
In the matrix element $T_1$, the only non-zero contribution comes from the term $G_0$ in Eq.~\eqref{Hrad},
	\begin{align}\label{T1G0}
		T_1=i\dfrac{\sqrt{3}}{4\pi} \int d^3rR_0\,G_0\,R_1\,(\bm e_\mu\cdot\bm n)\,,\quad\mu=J_z\,,
	\end{align}	 
where the following relations are  used
	\begin{align}
		&Y_{1\,\mu}=i\,\sqrt{\dfrac{3}{4\pi}}(\bm e_\mu\cdot\bm n)\,,\quad Y_{00}=\dfrac{1}{\sqrt{4\pi}} \,,\quad \bm n=\bm r/r\,,\nonumber\\
		&\bm e_0=\bm e_z\,,\quad \bm e_1=-\dfrac{(\bm e_x+i\bm e_y)}{\sqrt{2}}\,,\quad \bm e_{-1}=\dfrac{(\bm e_x-i\bm e_y)}{\sqrt{2}}\,.
	\end{align} 
In $T_2$, the only non-zero contribution comes from the term $\bm \Sigma\cdot\bm G_{\Sigma}/2$ in Eq.~\eqref{Hrad}. After simple transformations, we arrive at the result:
	\begin{align} \label{T2GS}
		& T_2= \dfrac{\sqrt{3}}{8\pi \sqrt{2}}\int d^3r\,R_0\,\Big(\bm e_\mu\cdot[\bm G_{\Sigma}\times\bm n]\Big)\, R_1\,,
	\end{align}
Using the obtained wave functions, we find
	\begin{align}\label{T1T2Z}
		&T_1=Z\,(\bm A_W\cdot\bm e_\mu)\,t_1\,,\quad T_2=Z\,(\bm A_W\cdot\bm e_\mu)\,t_2\,,\nonumber\\
		& Z=\dfrac{\omega_0\omega_1^2\sqrt{\omega_0\omega_1}}{\bar\omega^4}\,,\quad \bar\omega=\sqrt{\dfrac{\omega_0^2+\omega_1^2}{2}}\,.
	\end{align}
	For the coefficient $t_1$, the straightforward calculation gives $t_{1}=t_{11}+t_{12}$, where
	\begin{align} \label{t1}
		& t_{11}=	-\dfrac{\omega_0^2}{\sqrt{2}\,\bar\omega}
		\left(\dfrac{e_{\bar s}}{m_s}\,e^{-Q_s^2/4}-\dfrac{e_c}{m_c}\,e^{-Q_c^2/4}\right)\,,\nonumber\\
		&t_{12}=	\dfrac{5\omega_0^4\,\omega_1^2}{4\sqrt{2}\,\bar\omega^3}\,\Bigg\{\dfrac{e_{\bar s}}{m_s^3}
		\,e^{-Q_s^2/4}\left[1+\dfrac{Q_s^2(\omega_0^4+\omega_1^4)}{20\omega_0^2\omega_1^2}\right]-\dfrac{e_{c}}{m_c^3}
		\,e^{-Q_c^2/4}\left[1+\dfrac{Q_c^2(\omega_0^4+\omega_1^4)}{20\omega_0^2\omega_1^2}\right]\Bigg\}\,,\nonumber\\
        &Q_s=\dfrac{M_Rk}{m_s\,\bar\omega}\,,\quad Q_c=\dfrac{M_Rk}{m_c\,\bar\omega}\,.
		\end{align}
Note that  account for retardation, i.e.  $e^{-i\bm q_{\bar s}\cdot\bm r}$ and $e^{i\bm q_{c}\cdot\bm r}$ in the photon wave function, and compensation in the difference $e_{\bar s}/m_s-e_{c}/m_c$ leads to a decrease in the value of nonrelativistic contribution $t_{11}$ by half. Therefore, it is important to take into account the relativistic corrections $t_{12}$ and $t_2$. For $t_2$ we find $t_2=t_{21}+t_{22}+t_{23}$, where
	\begin{align} \label{t2}
		&t_{21}=\dfrac{1}{\sqrt{\pi}}\int_0^\infty re^{-r^2}
		\Bigg\{e_{\bar s}\left[\dfrac{(br^2-g\,\bar\omega^2)}{m_s^2}+\dfrac{2g\,\bar\omega^2}{m_sm_c}\right]\,{\cal F}(Q_sr)\nonumber\\
		&+e_{c}\left[\dfrac{(br^2-g\,\bar\omega^2)}{m_c^2}+\dfrac{2g\,\bar\omega^2}{m_sm_c}\right]\,{\cal F}(Q_cr)\Bigg\}\,dr\,,\nonumber\\
		& {\cal F}(x)=\dfrac{\sin x}{x}+\dfrac{\cos x}{x^2}-\dfrac{\sin x}{x^3}\,,\nonumber\\
		&t_{22}=\dfrac{k}{4}
		\left(\dfrac{e_{\bar s}}{m_s}Q_s\,e^{-Q_s^2/4}+\dfrac{e_{c}}{m_c}Q_c\,e^{-Q_c^2/4}\right)\,,\nonumber\\
		&t_{23}=\dfrac{k}{16\, \bar\omega}\left[4\omega_0^2
		\left(\dfrac{e_{\bar s}}{m_s^2}\,e^{-Q_s^2/4}+\dfrac{e_{c}}{m_c^2}\,e^{-Q_c^2/4}\right)+\bar\omega^2\left(\dfrac{e_{\bar s}}{m_s^2}Q_s^2\,e^{-Q_s^2/4}+\dfrac{e_{c}}{m_c^2}Q_c^2\,e^{-Q_c^2/4}\right)\right]\,.
	\end{align}	
As a result, we obtain  the transition probabilities,
	\begin{align} \label{G12}
		& \Gamma(D_{s1}(2460)\to D_s\gamma)=\dfrac{4}{3}k_1Z^2(c_1 t_1+c_2t_2)^2\,,\nonumber\\
		&\Gamma(D_{s1}(2536)\to D_s\gamma) =	\dfrac{4}{3}k_2Z^2(c_2 t_1-c_1t_2)^2\,.
	\end{align} 
When calculating these quantities, it is necessary to substitute the corresponding photon energies,
	\begin{align} \label{k12}
		&k_1=k_{10}-\dfrac{k_{10}^2}{2M_0}\,,\quad k_2=k_{20}-\dfrac{k_{20}^2}{2M_0}\,,\nonumber\\
		& k_{10}=[E_1^{(0)}+\Delta E_1+(\Delta H_{nr})_{11}]-[E_0^{(0)}+(\Delta H_{nr})_{00}+(\Delta H_{S})_{00}]\,,\nonumber\\
		&k_{20}=k_1+\Delta E_2-\Delta E_1\,.
	\end{align} 
	In these formulas  the recoil energies are taken into account, $M_0$ is the mass of $D_s$ meson.
	Explicit expressions for all the quantities included here are given in Eqs.~\eqref{c12},\eqref{E00},\eqref{E10}, \eqref{H12H22}, \eqref{delnr1}, \eqref{delnr0}, and \eqref{delS0}.
	
	For numerical estimates we use the  values of parameters
	$$m_s= 0.5\,\mbox{GeV}\,,\quad m_c= 1.7\,\mbox{GeV}\,,\quad b=0.18\,\mbox{GeV}^2\,,\quad g=0.8\,.$$
	With these parameters we have
	\begin{align} \label{ExpF}
				&\omega_0=0.47\,\mbox{GeV}\,,\quad \omega_1=0.38\,\mbox{GeV}\,,\nonumber\\
		&c_1=0.71\,,\quad c_2=0.71\,,\quad \varphi=\arctan(c_2/c_1)=45^\circ\,.
	\end{align}
	For the transitions $D_{s1}(2460)\to D_s\gamma$ and  $D_{s1}(2536)\to D_s\gamma$  we find from \eqref{G12}
	\begin{align} \label{G1F}
				&\Gamma(D_{s1}(2460)\to D_s\gamma)=297\,\mbox{KeV}\,,\nonumber\\
&\Gamma(D_{s1}(2536)\to D_s\gamma) =12\,\mbox{KeV}\,.
                	\end{align}
This result answers the question of \cite{Bondar:2025gsh}, why the decay $D_{s1}(2536)\to D_s\gamma$ is not observed, in contrast to the decay $D_{s1}(2460)\to D_s\gamma$.

	\subsection {Radiative transitions of $D_{s1}$ mesons to $D_{s}^*\gamma$}
The state of  $D_s^*$ meson is described by the wave function $\psi_0'$,
	\begin{align}
		\psi_0'=Y_{0\,0}\,|S=1,S_z=\mu'\rangle\,R_0(r)\,.
	\end{align}
	We need to calculate two matrix elements,
	\begin{align}\label{T1T2S}
		T_1'=\langle \psi_0'|H_{rad}|\psi_1\rangle\,, \quad  T_2'=\langle \psi_0'|H_{rad}|\psi_2\rangle\,.
	\end{align}	 
 The non-zero contribution to the matrix element $T_1'$ comes from the term $\bm \Sigma\cdot\bm G_{\Sigma}/2$ in Eq.~\eqref{Hrad},
	\begin{align}\label{T1P}
		T_1'=i\dfrac{\sqrt{3}}{8\pi} \int d^3rR_0\,(\bm G_\Sigma\cdot\bm e^*_{\mu'})(\bm n\cdot\bm e_\mu)\,R_1\,.
	\end{align}	 
The non-zero contributions to  $T_2'$ come from the terms $G_0$ and $\bm S_{tot}\cdot\bm G_{tot}/2$,
	\begin{align}\label{T2P}
		&	T_2'=-\dfrac{\sqrt{3}}{4\pi\sqrt{2}} \int d^3rR_0\,G_0\,([\bm e^*_{\mu'}\times\bm e_\mu]\cdot\bm n)\,R_1\,,\nonumber\\
		&-i\dfrac{\sqrt{3}}{8\pi\sqrt{2}} \int d^3rR_0\,([\bm G_{tot}\times\bm e^*_{\mu'}]\cdot[\bm n\times\bm e_\mu])\,R_1\,.
	\end{align}
Calculating the matrix elements yields the following expression for the amplitude $T_1'$:
	\begin{align}\label{T1PF}
		T_1'=iZ\Big\{ (\bm A_W\cdot[\bm e^*_{\mu'}\times\bm e_\mu]) \,\tau_{11} +   (\bm A_W\cdot[\bm e^*_{\mu'}\times\bm \lambda])(\bm e_\mu\cdot\bm\lambda )\,\tau_{12}\Big\}\,,
	\end{align}
where $\bm\lambda=\bm k/k$, $Z$ is defined in \eqref{T1T2Z}, the coefficients $\tau_{11}$ and $\tau_{12}$ are 
		\begin{align} \label{tau11}
		&\tau_{11}=\dfrac{\sqrt{2}}{\sqrt{\pi}}\int_0^\infty re^{-r^2}
		\Bigg\{e_{\bar s}\left[\dfrac{(br^2-g\,\bar\omega^2)}{m_s^2}+\dfrac{2g\,\bar\omega^2}{m_sm_c}\right]\,{\cal F}_1(Q_sr)\nonumber\\
		&+e_{c}\left[\dfrac{(br^2-g\,\bar\omega^2)}{m_c^2}+\dfrac{2g\,\bar\omega^2}{m_sm_c}\right]\,{\cal F}_1(Q_cr)\Bigg\}\,dr
		+\dfrac{k\omega_0^2}{4\sqrt{2}\, \bar\omega}
		\left(\dfrac{e_{\bar s}}{m_s^2}\,e^{-Q_s^2/4}+\dfrac{e_{c}}{m_c^2}\,e^{-Q_c^2/4}\right)\,,\nonumber\\
		& {\cal F}_1(x)=\dfrac{\sin x}{x^3}-\dfrac{\cos x}{x^2}\,,
	\end{align}	
	\begin{align} \label{tau12}
		&\tau_{12}=\dfrac{\sqrt{2}}{\sqrt{\pi}}\int_0^\infty re^{-r^2}
		\Bigg\{e_{\bar s}\left[\dfrac{(br^2-g\,\bar\omega^2)}{m_s^2}+\dfrac{2g\,\bar\omega^2}{m_sm_c}\right]\,{\cal F}_2(Q_sr)\nonumber\\
		&+e_{c}\left[\dfrac{(br^2-g\,\bar\omega^2)}{m_c^2}+\dfrac{2g\,\bar\omega^2}{m_sm_c}\right]\,{\cal F}_2(Q_cr)\Bigg\}\,dr
		+\dfrac{M_Rk^2}{2\sqrt{2}\, \bar\omega}
		\left(\dfrac{e_{\bar s}}{m_s^2}\,e^{-Q_s^2/4}+\dfrac{e_{c}}{m_c^2}\,e^{-Q_c^2/4}\right)\nonumber\\
		&+\dfrac{M_R^2k^3}{8\sqrt{2}\, \bar\omega}\left(1-\dfrac{\omega_0^2}{\bar\omega^2}\right)
		\left(\dfrac{e_{\bar s}}{m_s^4}\,e^{-Q_s^2/4}+\dfrac{e_{c}}{m_c^4}\,e^{-Q_c^2/4}\right)\,,\nonumber\\
		& {\cal F}_2(x)=\dfrac{\sin x}{x}-\dfrac{3\sin x}{x^3}+\dfrac{3\cos x}{x^2}\,.
	\end{align}
The amplitude $T_2'$ reads
	\begin{align}\label{T2PF}
		T_2'=iZ\Big\{ (\bm A_W\cdot[\bm e^*_{\mu'}\times\bm e_\mu]) \,\tau_{21} +   (\bm A_W\cdot[\bm e_\mu \times\bm \lambda])(\bm e^*_{\mu'}\cdot\bm\lambda )\,\tau_{22}\Big\}\,,
	\end{align}
where the coefficients $\tau_{21}$ and $\tau_{22}$ are 
	\begin{align} \label{tau11}
		&\tau_{21}=\dfrac{1}{\sqrt{\pi}}\int_0^\infty re^{-r^2}
		\Bigg\{e_{\bar s}\left[-\dfrac{(br^2-g\,\bar\omega^2)}{m_s^2}+\dfrac{2g\,\bar\omega^2}{m_sm_c}\right]\,{\cal F}_1(Q_sr)\nonumber\\
		&-e_{c}\left[-\dfrac{(br^2-g\,\bar\omega^2)}{m_c^2}+\dfrac{2g\,\bar\omega^2}{m_sm_c}\right]\,{\cal F}_1(Q_cr)\Bigg\}\,dr
		-\dfrac{k\omega_0^2}{8\, \bar\omega}
		\left(\dfrac{e_{\bar s}}{m_s^2}\,e^{-Q_s^2/4}-\dfrac{e_{c}}{m_c^2}\,e^{-Q_c^2/4}\right)\nonumber\\
		&	-\dfrac{\omega_0^2}{2\bar\omega}
		\left(\dfrac{e_{\bar s}}{m_s}\,e^{-Q_s^2/4}-\dfrac{e_{c}}{m_c}\,e^{-Q_c^2/4}\right)+
		\dfrac{5\omega_0^4\,\omega_1^2}{8\,\bar\omega^3m_s^3}\,e_{\bar s}
		\,e^{-Q_s^2/4}\left[1+\dfrac{Q_s^2(\omega_0^4+\omega_1^4)}{20\omega_0^2\omega_1^2}\right] \nonumber\\
        &-\dfrac{5\omega_0^4\,\omega_1^2}{8\,\bar\omega^3m_c^3}\,e_{c}
		\,e^{-Q_c^2/4}\left[1+\dfrac{Q_c^2(\omega_0^4+\omega_1^4)}{20\omega_0^2\omega_1^2}\right]\,,
	\end{align}	
	\begin{align} \label{tau21}
		&\tau_{22}=-\dfrac{1}{\sqrt{\pi}}\int_0^\infty re^{-r^2}
		\Bigg\{e_{\bar s}\left[-\dfrac{(br^2-g\,\bar\omega^2)}{m_s^2}+\dfrac{2g\,\bar\omega^2}{m_sm_c}\right]\,{\cal F}_2(Q_sr)\nonumber\\
		&-e_{c}\left[-\dfrac{(br^2-g\,\bar\omega^2)}{m_c^2}+\dfrac{2g\,\bar\omega^2}{m_sm_c}\right]\,{\cal F}_2(Q_cr)\Bigg\}\,dr
		+\dfrac{M_Rk^2}{4\, \bar\omega}
		\left(\dfrac{e_{\bar s}}{m_s^2}\,e^{-Q_s^2/4}-\dfrac{e_{c}}{m_c^2}\,e^{-Q_c^2/4}\right)\nonumber\\
		&+\dfrac{M_Rk^2}{16}\left(1-\dfrac{\omega_0^2}{\bar\omega^2}\right)
		\left(\dfrac{e_{\bar s}}{m_s^3}\,e^{-Q_s^2/4}-\dfrac{e_{c}}{m_c^3}\,e^{-Q_c^2/4}\right)\,.
	\end{align}
The probabilities of radiative transitions are of the form
	\begin{align} \label{G12S}
		& \Gamma(D_{s1}(2460)\to D_s^*\gamma)=\dfrac{4}{3}k_1Z^2\Big\{[c_1\tau_{11}+c_2(\tau_{21}-\tau_{22})]^2+ [c_1(\tau_{11}+\tau_{12})+c_2\tau_{21}]^2\Big\}\,,\nonumber\\
		&\Gamma(D_{s1}(2536)\to D_s^*\gamma) =\dfrac{4}{3}k_2Z^2\Big\{[-c_2\tau_{11}+c_1(\tau_{21}-\tau_{22})]^2+ [-c_2(\tau_{11}+\tau_{12})+c_1\tau_{21}]^2\Big\}\,.
	\end{align}
The photon momenta in radiative transitions are given in Eq.~\eqref{k12} with $(\Delta H)_{00}$ replaced by
		\begin{align}
		(\Delta H)_{01}=\dfrac{2g\, \omega_0^3 }{3\sqrt{\pi}\, m_sm_c}\,.
	\end{align}
For numerical estimates, we used the same parameters as before. As a result we obtain
	\begin{align} \label{ExpFDSS}
				&\Gamma(D_{s1}(2460)\to D_s^*\gamma)= 104\,\mbox{KeV}\,, \nonumber\\
		&\Gamma(D_{s1}(2536)\to D_s^*\gamma) = 29\,\mbox{KeV}\,.
	\end{align}
	
	\subsection{Comparison with previous predictions}
	
	\begin{table}[!h]
		\caption{\label{tb1} Theoretical predictions for radiative decay widths $D_{s1}\to D_s^{(*)}\gamma$ in \mbox{KeV}.
		}
		\begin{center}
			
			%\begin{ruledtabular}
			\begin{tabular}{lrrrrrrrrr} 
				& \multicolumn{1}{c}{ }&\multicolumn{1}{c}{ }&\multicolumn{1}{c}{ }&\multicolumn{1}{c}{ }&\multicolumn{1}{c}{ }&\multicolumn{1}{c}{ }&\multicolumn{1}{c}{ }&\multicolumn{1}{c}{ }&\multicolumn{1}{c}{ }\\

				Publication & \multicolumn{1}{c}{ \cite{Godfrey:2005ww} }  &\multicolumn{1}{c}{ \cite{Goity:2000dk} } &   &\multicolumn{1}{c}{ \cite{Green:2016occ}   } &\multicolumn{1}{c}{\cite{Radford:2009bs}  } &\multicolumn{1}{c}{ \cite{Chen:2020jku} } & &\multicolumn{1}{c}{ \cite{Korner:1992pz}  } &\multicolumn{1}{c}{This work }\\
				\hline
				
				$\Gamma(\DsTS \to D_s\gamma) $  & 15 &25.2-31.1 &  & 61.2 & 54.5 & 18.18-18.85 &  & 1.6 $\pm$ 2.3 & 12\\
				$\Gamma(\DsTS \to D_s^{*} \gamma) $  & 5.6 & 14.6-22.8 &  & 9.21 & 8.9 & 2.96-3.02 &  & 10.4 $\pm$ 1. & 29\\
				$\Gamma(\DsFE \to D_s \gamma) $  & 6.2 & 10.3-17.2 &  & 13.2 & 12.8 & 3.53-3.61 &  &  & 297\\ 
				$\Gamma(\DsFE \to D_s^{*} \gamma) $  & 5.5 & 14.0-25.1 &  & 17.4 & 15.5 & 4.74-4.79 &  &  & 104\\
				\hline
				
			\end{tabular}
			%\end{ruledtabular}
			
		\end{center}
	\end{table}
	
	Numerous theoretical works \cite{Godfrey:2005ww,Goity:2000dk,Green:2016occ,Radford:2009bs,Chen:2020jku,Korner:1992pz} are devoted to  calculation of radiative decay partial widths of $D_{s1}$ mesons. The results of these studies, carried out within the framework of potential models, are presented in Table \ref{tb1} together with the predictions of our work. It is evident that in all previous studies the predictions for the width $\Gamma(\DsTS \to D_s \gamma) $ are larger than that for $\Gamma(\DsFE \to D_s \gamma) $, and the decay width $\Gamma(\DsFE \to D_s^{*} \gamma) $ systematically exceeds $\Gamma(\DsFE \to D_s \gamma) $, which significantly contradicts the experimental data \cite{ParticleDataGroup:2024cfk}.
	Apparently, this is due to the fact that in all previous works the spin-dependent contributions to the radiative transition amplitudes have been neglected. Therefore, the ratio of transition probabilities to $D_s$ is determined only by the mixing angle, which is of the order of forty degrees. As can be seen from our calculations,  the spin-independent contribution to the transition amplitudes is strongly suppressed.
	As a result, the contributions of relativistic corrections, dependent on quark spins and independent of them, become very important. As it turned out, the  relativistic corrections to the amplitudes of radiative transitions is of the same order as the contribution of nonrelativistic dipole moment operator.
	Since in the decay amplitude $\DsTS \to D_s \gamma $ the corresponding contributions enter with opposite signs \eqref{G12S}, a strong compensation occurs in the corresponding decay width. At the same time, an amplification takes place in the decay amplitude $\DsFE \to D_s \gamma $. A similar phenomenon occurs in transitions to $D_s^*\gamma$.
	
Of course, our predictions are model-dependent, and the numerical values of widths are sensitive to such parameters as the quark masses and the constants in the interaction potential. However, at our values of the parameters, we reproduce well the experimentally known mass differences of $D_{s1}$, $D_s$ and $D_s^{*}$ mesons.
The sensitivity of radiative widths to small changes in model parameters is more significant. 
Since $\Gamma(\DsTS \to D_s \gamma) $ turned out to be very small as a result of strong compensation, we believe that this value lies in the range from zero to $15\,\mbox{KeV}$.
At the same time, we estimate the accuracy of the remaining predictions at $50\%$. Note that such a variation of widths does not significantly affect their ratios. 
Our statement, that the decay width of $\DsFE \to D_s \gamma $ transition is much larger than that for $\DsTS \to D_s \gamma $, is stable with respect to variations in the model parameters. 

Since the decay probability of $\DsFE \to D_s \gamma $ is known from experiment, we can estimate the total width of $\DsFE$ meson, $\Gamma_{tot}(\DsFE)\sim 1.5\,\mbox{MeV}$. Therefore, measuring the total width of $\DsFE$ is an important check of our estimates. In addition, measuring the ratio $\Gamma(D_{s1}(2460)\to D_s^*\gamma)/\Gamma(D_{s1}(2460)\to D_s\gamma) $ will also be a test of the model. This ratio is much less dependent on the choice of parameters than the partial widths and is easier to measure.
 
	\section{Radiative transitions of $B_{s1}$ and $B_{c1}$ mesons}
	Using the obtained formulas, we can predict the partial widths for radiative transitions
	 $B_{s1}\to B_{s}\gamma$, $B_{s1}\to B_{s}^*\gamma$, $B_{c1}\to B_{c}\gamma$ and $B_{c1}\to B_{c}^*\gamma$.
	
	\subsection {Radiative transitions $B_{s1}\to B_{s}\gamma$ and $B_{s1}\to B_{s}^*\gamma$} 
	Let us make the substitution $m_c\to m_b=4.8\,\mbox{GeV}$ and $e_c\to e_b=-e/3$ in the formulas obtained for $D_{s1}$ mesons. In addition, for $B_{s1}$ mesons we use the value $g=0.7\,$. The small decrease in the constant $g$ is due to the fact that with the growth of  heavy quark mass  the reduced mass $M_R$ increases, which leads to a decrease in the characteristic size of the system.  As a result, the strong interaction coupling constant $\alpha_s$ and, consequently, the constant $g$ decrease. Then we have
	\begin{align} \label{ExpFB}
		&\omega_0=0.49\,\mbox{GeV}\,,\quad \omega_1=0.40\,\mbox{GeV}\,,\nonumber\\
		&c_1=0.78\,,\quad c_2=0.62\,,\quad \varphi=\arctan(c_2/c_1)=38.3^\circ\,.
	\end{align}
	For the transition $B_{s1}(5748)\to B_s\gamma$ and $B_{s1}(5829)\to B_s\gamma$ we obtain
		\begin{align} \label{G1BF}
				&\Gamma(B_{s1}(5748)\to B_s\gamma)=47\,\mbox{KeV}\,,\nonumber\\
		&\Gamma(B_{s1}(5829)\to B_s\gamma) =28\,\mbox{KeV}\,.
	\end{align}
For the transitions of $B_{s1}$ mesons to $B_{s}^*\gamma$ we find
		\begin{align} \label{ExpFBSS}
		&\Gamma(B_{s1}(5748)\to B_s^*\gamma)=20\,\mbox{KeV}\,, \nonumber\\
		&\Gamma(B_{s1}(5829)\to B_s^*\gamma) =41\,\mbox{KeV}\,.
	\end{align}

A comparison of our results with the predictions of other authors is given in Table~ \ref{tb2}.
    
    \begin{table}[!h]
\caption{\label{tb2} Theoretical predictions for radiative decay widths  $B_{s1}\to B_s^{(*)}\gamma$ in \mbox{KeV}.
}
\begin{center}

%\begin{ruledtabular}
\begin{tabular}{lrrrrrrrrr} 
    & \multicolumn{1}{c}{ }&\multicolumn{1}{c}{ }&\multicolumn{1}{c}{ }&\multicolumn{1}{c}{ }&\multicolumn{1}{c}{ }&\multicolumn{1}{c}{ }&\multicolumn{1}{c}{ }&\multicolumn{1}{c}{ }&\multicolumn{1}{c}{ }\\

 Publication & \multicolumn{1}{c}{ }  &\multicolumn{1}{c}{  } &\multicolumn{1}{c}{\cite{li:2021hss}  } &\multicolumn{1}{c}{ \cite{Godfrey:2016nwn}   } &\multicolumn{1}{c}{\cite{Lu:2016bbk}  } &\multicolumn{1}{c}{  } &\multicolumn{1}{c}{  } &\multicolumn{1}{c}{   } &\multicolumn{1}{c}{This work      }\\
 \hline

$\Gamma(B_{s1}(5748)\to B_s\gamma) $  &  & & 37 & 70.6 & 97.7 &  &  &  & 47\\
$\Gamma(B_{s1}(5748)\to B_s^*\gamma) $  &  &  & 56 & 36.9 & 39.5 &  &  &  & 20\\
$\Gamma(B_{s1}(5829)\to B_s\gamma) $  &  &  & 27 & 47.8 & 56.6 &  &  &  & 28\\ 
$\Gamma(B_{s1}(5829)\to B_s^*\gamma) $  &  &  & 53 & 57.3 & 98.8 &  &  &  & 41\\
\hline
          
\end{tabular}
%\end{ruledtabular}

\end{center}
\end{table}

	%%%%%%%%%%%%%%%%%
	
	\subsection {Radiative transitions $B_{c1}\to B_{c}\gamma$ and $B_{c1}\to B_{c}^*\gamma$
	} 
The expressions for probabilities of radiative transitions $B_{c1}\to B_{c}\gamma$ and $B_{c1}\to B_{c}^*\gamma$ are obtained from that for  transitions $D_{s1}\to D_{s}\gamma$ and $D_{s1}\to D_{s}^*\gamma$ by substituting $m_c\to m_b=4.8\,\mbox{GeV}$, $m_s\to m_c=1.7\,\mbox{GeV}$, $e_c\to e_b=-e/3$ and $e_{\bar s}\to e_{\bar c}=-2e/3$. In addition, we use the value $g=0.6$ for these transitions. As a result we have 
	\begin{align} \label{ExpFBc}
		&\omega_0=0.82\,\mbox{GeV}\,,\quad \omega_1=0.60\,\mbox{GeV}\,,\nonumber\\
		&c_1=-0.22\,,\quad c_2=0.98\,,\quad \varphi=\arctan(c_2/c_1)=-77^\circ\,.
	\end{align}
For  transitions $B_{c1}(6743)\to B_c\gamma$ and $B_{c1}(6750)\to B_c\gamma$ we obtain
	\begin{align} \label{G1BFBc}
		&\Gamma(B_{c1}(6743)\to B_c\gamma)=22\,\mbox{KeV}\,\nonumber\\
		&\Gamma(B_{c1}(6750)\to B_c\gamma) =47\,\mbox{KeV}\,.    \end{align}
		Interestingly, for $B_{c1}\to B_{c}\gamma$ transition, the probability of emission from the lower level is very small, compared to the probability of emission from the upper level.
	
	For  transitions $B_{c1}\to B_{c}^*\gamma$ we find
		\begin{align} \label{ExpFBCS}
				&\Gamma(B_{c1}(6743)\to B_c^*\gamma)=75\,\mbox{KeV}\,, \nonumber\\
		&\Gamma(B_{c1}(6750)\to B_c^*\gamma) =2\,\mbox{KeV}\,.
	\end{align}

A comparison of our results with the predictions of other authors is given in Table~\ref{tb3}.

    %\cite{Godfrey:2004ya,Ebert:2002pp,Fulcher:1998ka,Gershtein:1994dxw,Gupta:1995ps,Wang:2022cxy,Li:2023wgq,Hao:2024nqb,Li:2019tbn,Asghar:2019qjl,Li:2022bre,Eichten:2019gig}.

    \begin{table}[!h]
\caption{\label{tb3} Theoretical predictions for radiative decay widths $B_{c1}\to B_c^{(*)}\gamma$ in \mbox{KeV}.
}
\begin{center}

%\begin{ruledtabular}
\begin{tabular}{lrrrrrrrrr} 
    & \multicolumn{1}{c}{ }&\multicolumn{1}{c}{ }&\multicolumn{1}{c}{ }&\multicolumn{1}{c}{ }&\multicolumn{1}{c}{ }&\multicolumn{1}{c}{ }&\multicolumn{1}{c}{ }&\multicolumn{1}{c}{ }&\multicolumn{1}{c}{ }\\

 Publication & \multicolumn{1}{c}{ \cite{Godfrey:2004ya} }  &\multicolumn{1}{c}{ \cite{Ebert:2002pp} } &\multicolumn{1}{c}{\cite{Fulcher:1998ka}  } &\multicolumn{1}{c}{ \cite{Gershtein:1994dxw}   } &\multicolumn{1}{c}{\cite{Li:2022bre}  } &\multicolumn{1}{c}{ \cite{Li:2019tbn} } &\multicolumn{1}{c}{\cite{Li:2023wgq}  } &\multicolumn{1}{c}{ \cite{Eichten:2019gig}  } &\multicolumn{1}{c}{This work       }\\
 \hline

$\Gamma(B_{c1}(6743)\to B_c\gamma) $  & 13 & 18.4 & 32.4 & 11.6 & 16.6 & 35 & 30.1 & 9.9 & 22\\
$\Gamma(B_{c1}(6743)\to B_c^*\gamma) $  & 60 & 78.9 & 75.6 & 77.8 & 49 & 70 & 47.8 & 62.5 & 75\\
$\Gamma(B_{c1}(6750)\to B_c\gamma) $  & 80 & 132 & 127.8 & 131.1 & 66.6 & 74 & 64 & 92.3 & 47\\ 
$\Gamma(B_{c1}(6750)\to B_c^*\gamma) $  & 11 & 13.6 & 26.2 & 8.1 & 10.5 & 40 & 25.6 & 7.5 & 2\\
\hline
          
\end{tabular}
%\end{ruledtabular}

\end{center}
\end{table}
		%%%%%%%%%%%%%%%%%%%%%
Unfortunately, there are currently no experimental data for radiative decays of  $B_{s1}$ and $B_{c1}$ mesons. 

Let's make a remark on the mixing angle $\varphi=\arctan(c_2/c_1)$. If we let the mass of a heavy quark go to infinity, then $\varphi$ will be either $-\arctan\sqrt{2}=-54.7^\circ$  or $\arctan(1/\sqrt{2})=35.3^\circ$ \cite{Matsuki:2010zy} depending on the ratio of the parameters $g$ and $b$. For $g=0.5$ and $b=0$ it will be $-54.7^\circ$, and for $g=0$ and $b=0.18$ it will be $35.3^\circ$. Therefore, it is seen that for all the cases considered, the heavy-quark approximation works rather good only for the system $b\bar s$ where  $m_b\gg m_s$.

	\section {Conclusion}
	Within the  phenomenological approach, predictions are obtained for the probabilities of  radiative transitions $D_{s1}\rightarrow D_s\gamma$, $D_{s1}\rightarrow D_s^*\gamma$, $B_{s1}\rightarrow B_s\gamma$, $B_{s1}\rightarrow B_s^*\gamma$, $B_{c1}\rightarrow B_c\gamma$ and $B_{c1}\rightarrow B_c^*\gamma$.
	These predictions explain the unusual ratio between the widths of different radiative transitions of $D_{s1}$ mesons noted in \cite{Bondar:2025gsh}.
	The effect appears due to a large suppression of nonrelativistic  contribution from the  electric dipole moment of $c\bar s$  system to the decay  amplitude. As a result, the  relativistic corrections to the radiation amplitude turned out to be dominant and the amplitudes \eqref{T1T2}  become identical in sign and close in magnitude.
	Since the  states $\Psi_{1,2}$ corresponding to  $D_{s1}$ mesons are superposition of  $\psi_{1,2}$ states with the coefficients $c_{1,2}$ close in value, a large compensation of contributions occurs in the transition amplitude $\Gamma(D_{s1}(2536)\to D_s\gamma)$. As a result, the width  $\Gamma(D_{s1}(2536)\to D_s\gamma)$  becomes strongly suppressed.
	
	All obtained predictions are in qualitative agreement with the available experimental results and constraints. All predictions are obtained with one standard set of parameters under natural variation of the coupling constant $g$. Our predictions can be tested in future measurements at BESIII, BELLE-II and LHCb collaborations.

	%%%%%%%%%%%%%%%%%%%%%%%%%%%%%%%%%%%%%%%%%%%%%%%%%%%%%%%%%%%%%%%%%%%%%%%%%%
\bibliographystyle{JHEP}
\bibliography{note}
	
\end{document}